\documentclass[aps,pra,reprint,showpacs,superscriptaddress,notitlepage,twocolumn]{revtex4-1}%
\usepackage[utf8]{inputenc}
\usepackage{inputenc}
\usepackage{amsfonts}
\usepackage{mathrsfs}
\usepackage{amsmath}
\usepackage{amssymb}
\usepackage{graphicx}
\usepackage{color}%
\usepackage{mathtools}
\usepackage{booktabs}
\usepackage{mleftright}
\usepackage[colorlinks,linkcolor=blue,citecolor=blue,urlcolor=blue,hyperindex,bookmarks=false,pdfstartview=FitH]{hyperref}
\date{\today}
\begin{document}

\title{Enantio-specific state transfer of chiral molecules
through enantio-selective shortcut-to-adiabaticity paths}
\author{Jian-Jian Cheng}
\affiliation{Center for Theoretical Physics and School of Science, Hainan University, Haikou 570228, China}
\affiliation{Beijing Computational Science Research Center, Beijing 100193, China}
\author{Chong Ye}
\affiliation{Beijing Key Laboratory of Nanophotonics and Ultrafine Optoelectronic Systems, School of Physics, Beijing Institute of Technology, 100081 Beijing, China}
\author{Yong Li}
\email{yongli@hainanu.edu.cn}
\affiliation{Center for Theoretical Physics and School of Science, Hainan University, Haikou 570228, China}
\affiliation{Synergetic Innovation Center for Quantum Effects and Applications, Hunan Normal University, Changsha 410081, China}
\begin{abstract}
An interesting method of fast enantio-specific state transfer is proposed for cyclic three-level systems of chiral molecules. We show that the fast population transfer via shortcut to adiabaticity can be accomplished for the cyclic three-level system of a general  (chiral) molecule with invariant-based inverse engineering of the coupling strengths. By choosing appropriate parameters, the two enantiomers, which are initially prepared in their ground states in the three-level systems, will evolve respectively along their enantio-selective shortcut-to-adiabaticity paths to different-energy final states simultaneously, namely achieving the fast enantio-specific state transfer.
\end{abstract}

\maketitle

\section{Introduction}
Since Pasteur first discovered chiral molecules in 1848, the theoretical and experimental studies of chiral molecules have proliferated in chemistry~\cite{chemical}, biotechnologies~\cite{biological}, and pharmaceutics~\cite{pharmaceutical}. Chiral molecules contain two species,
e.g. left- and right-handed ones~\cite{enantiomer},  which are often called enantiomers. The two enantiomers are mirror images of each other but can be superposed on each other via translations and rotations. The enantiodiscrimination (as well as enantioseparation and enantioconversion)~\cite{McKendry-nature1998-discriminiation,
Barron-nature2000-separation,Ahuja-book2011-separation,Zepik-science2002-conversion} of chiral molecules remains an enormous challenge.
The traditional method of enantiodiscrimination is to break the mirror symmetry of the enantiomers by using circularly polarized light~\cite{optical}.
Some commonly used chiroptical methods of enantiodiscrimination are circular dichroism~\cite{circular}, vibrating circular dichroism~\cite{vibrationl}, optical rotation~\cite{optical}, and Raman optical activity~\cite{activity}. However, these methods rely on the interference between electric-dipole and weak magnetic-dipole (or electric-quadrupole) transitions.

Alternatively, enantiodiscrimination methods that only use electric-dipole interactions~\cite{Shapiro-JCP1991-discrimination,Hirota-PJA2012-discrimination}, have also been proposed. The left- and right-handed chiral molecules can be modeled as cyclic three-level systems, where three electromagnetic (optical or microwave) fields couple respectively to three transitions via electric-dipole interactions~\cite{Kral-RPL2001-ESST,Kral-RPL2003-ESST}.
Due to the intrinsic property of chiral molecules, the product of the corresponding three coupling strengths (Rabi frequencies) in the cyclic three-level systems can differ in signs for the two enantiomers~\cite{Kral-RPL2001-ESST,Kral-RPL2003-ESST}. So the corresponding overall phases in the cyclic three-level systems differ by $\pi$ with the enantiomers. Based on such cyclic three-level systems, one can use different schemes, such as enantio-selective three-wave mixing~\cite{Patterson-Nature2013-discrimination,Patterson-PRL2013-discrimination,Shubert-Angew2014-discrimination,Lobsiger-JPCL2015-discrimination,
Shubert-JPCL2016-discrimination}, enantio-selective absorption~\cite{Jia-PRA2011-discrimination}, enantio-selective AC stark effect~\cite{Ye-PRA2019-discrimination}
and enantio-selective two-dimensional spectra~\cite{Cai-PRL2022-discrimination},
to discriminate the left- and right-handed molecules.
Moreover, some more ingenious sources of modern optics physics, such as frequency entangled photons~\cite{Ye-JPCL2021-discrimination}, quantized photons~\cite{Chen-OE2021-discrimination,Chen-PRR2022-discrimination}, and correlated photons in cavities~\cite{Zou-OE2021-discrimination},
have been introduced to enhance the performance of enantiodiscrimination.

Beyond the enantiodiscrimination, the cyclic three-level systems of chiral molecules have also been used in some more ambitious issues, such as the enantio-specific state transfer (ESST)~\cite{Kral-RPL2001-ESST,Li-PRA2007-ESST,Jia-JPB2010-ESST,Vitanov-PRL2019-ESST,Ye-PRA2019-ESST,Leibscher-JCP2019-ESST,Wu-PRA2019-ESST,Torosov-PRA2020-ESST,
Wu-PRApplied2020-ESST,Torosov-PRR2020-ESST,Zhang-JPB2020-ESST}, enantioseparation~\cite{Li-RRL2007-separation,Li-JCP2010-separation,Jacob-JCP2012-separation,Liu-PRA2021-separation}, and enantioconversion~\cite{Kral-RPL2003-ESST,Shapiro-PRL2000-conversion,Brumer-PRA65-conversion,Ye-PRR2020-conversion,Ye-PRA2021-conversion}. The perfect ESST of chiral molecules can be realized by transferring the left- and right-handed chiral molecules from the same-energy initial states to different-energy final states by choosing suitable electromagnetic fields~\cite{Kral-RPL2001-ESST,Li-PRA2007-ESST,Jia-JPB2010-ESST,Vitanov-PRL2019-ESST,Ye-PRA2019-ESST,Leibscher-JCP2019-ESST,Wu-PRA2019-ESST,Torosov-PRA2020-ESST,
Wu-PRApplied2020-ESST,Torosov-PRR2020-ESST,Zhang-JPB2020-ESST}. Recently, the feasibility of ESST based on the cyclic three-level systems has been demonstrated experimentally in gaseous samples by using microwave fields~\cite{Eibenberger-PRL2017-Experimental,Perez-Angew2017-Experimental,Perez-JPCL2018-Experimental,Lee-PRL2022-ESST}. After the achievement of the ESST, one can further  realize the enantiodiscrimination and spatial enantioseparation for the chiral molecules~\cite{Li-RRL2007-separation,Li-JCP2010-separation}.

In the original ESST method based on cyclic three-level systems of chiral molecules~\cite{Kral-RPL2001-ESST}, the ESST was realized by using the adiabatic (and also diabatic) passage technique, which makes the ESST process slow and complicated. To overcome these defects, several theoretical methods of fast ESST were proposed and developed~\cite{Li-PRA2007-ESST,Jia-JPB2010-ESST,Vitanov-PRL2019-ESST,Ye-PRA2019-ESST,Leibscher-JCP2019-ESST,Wu-PRA2019-ESST,Torosov-PRA2020-ESST,Torosov-PRR2020-ESST,
Wu-PRApplied2020-ESST,Zhang-JPB2020-ESST} based on cyclic three-level systems. Among them, an ingenious method~\cite{Vitanov-PRL2019-ESST} was proposed to achieve the fast ESST of chiral molecules by using the ``shortcut to adiabaticity'' (STA) concept via adding a counterdiabatic field to accelerate the stimulated Raman adiabatic passage.

Motivated by Ref.~\cite{Vitanov-PRL2019-ESST}, here we propose to achieve the ESST by a different STA with invariant-based inverse engineering~\cite{Chen-PRL2010-STA,Chen-PRA2011-STA,Chen-PRA2012-STA}, instead of the STA with adding the counterdiabatic field~\cite{Vitanov-PRL2019-ESST}.
The invariant-based inverse engineering starts by introducing a Lewis-Riesenfeld invariant in a time-dependent system. The invariant can be used to derive a law that governs the evolution state for the designed Hamiltonian. By means of the invariant-based inverse engineering of the time-dependent Hamiltonians with designing appropriate control parameters, the left- and right-handed chiral molecules prepared initially in their corresponding ground states would evolve (approximately) along their enantio-selective shortcut-to-adiabaticity paths to different-energy final states.

\section{CYCLIC THREE-LEVEL SYSTEMS}

\begin{figure}[htbp]
\centering
\includegraphics[width=250pt]{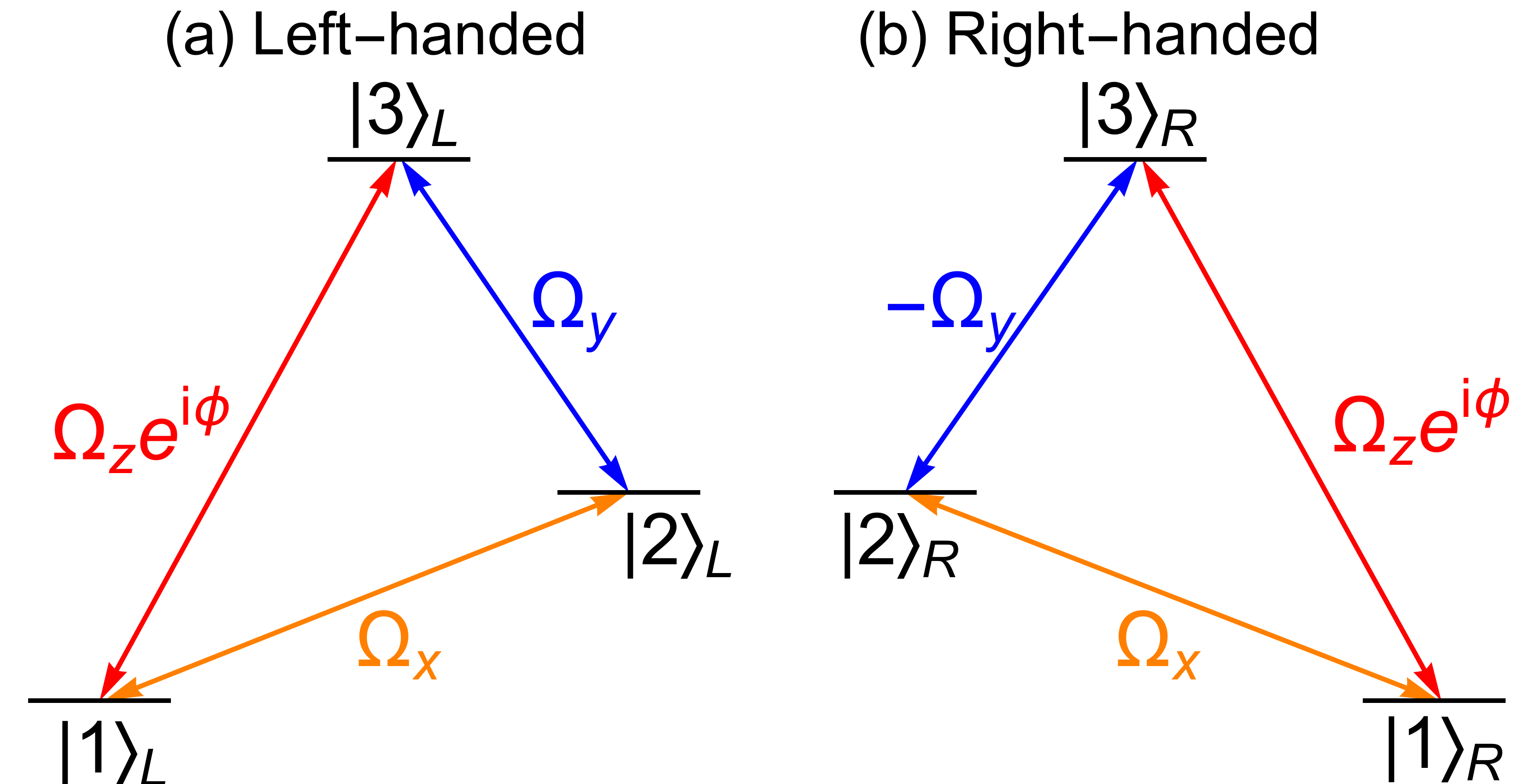}
\caption{(a) Left- and (b) right-handed chiral molecules of cyclic three-level systems, where three electromagnetic fields couple resonantly to the three electric-dipole transitions, respectively, with $\Omega_{x}$, $\pm\Omega_{y}$, and $\Omega_{z}e^{i\phi}$ the corresponding Rabi frequencies.}
\label{fig1}
\end{figure}

A general chiral molecule can be modeled as the cyclic three-level system by choosing appropriate three electromagnetic fields to couple with three electric-dipole transitions~\cite{Kral-RPL2001-ESST,Ye-PRA2018}. Here, we only consider the case that all the three electromagnetic fields couple resonantly with the electric-dipole transitions respectively, as shown similar to Fig.~\ref{fig1}(a). In the basis of $\{|1\rangle, |2\rangle, |3\rangle\}$, the Hamiltonian of the cyclic three-level system can be described in the interaction picture as $(\hbar=1)$~\cite{Vitanov-PRL2019-ESST}
\begin{eqnarray}
\hat{H}(t)=
\left(  \begin{array}{cccc}
0 & \Omega_{x}(t) & \Omega_{z}(t)e^{-i\phi}  \\
\Omega_{x}(t) & 0 & \Omega_{y}(t) \\
\Omega_{z}(t)e^{i\phi} & \Omega_{y}(t) & 0\\
  \end{array}
\right)
\label{1}
\end{eqnarray}
with $|1\rangle=(1,\,0,\,0)^{T},~|2\rangle=(0,\,1,\,0)^{T},~|3\rangle=(0,\,0,\,1)^{T}$. Here $\Omega_{j}(t)~(j=x,y,z)$ are the Rabi frequencies, which can be controlled by varying the amplitudes of the applied electromagnetic fields. $\phi$ is the overall phase of the three Rabi frequencies. Here we set $\phi=\pi/2$.
Without loss of generality, we have assumed $\Omega_{j}$ are real. Then the Hamiltonian can be expressed as
\begin{eqnarray}
\label{Hamilton}
\hat{H}(t)=\Omega_{x}(t)\hat{K}_{x}+\Omega_{y}(t)\hat{K}_{y}+\Omega_{z}(t)\hat{K}_{z}.
\end{eqnarray}
Here, $\hat{K}_{x}$, $\hat{K}_{y}$, and $\hat{K}_{z}$ are the SU(2) angular-momentum operators~\cite{Carroll-Opt1988-spin}
\begin{eqnarray}
\label{SU2}
\begin{aligned}
\hat{K}_{x}&=
\left(\begin{array}{cccc}
  0& 1 & 0  \\
1 & 0 & 0 \\
  0  & 0 & 0\\
  \end{array}
\right), \ \
\hat{K}_{y}=
\left(\begin{array}{cccc}
  0& 0 & 0  \\
0 & 0 & 1 \\
  0  & 1 & 0\\
  \end{array}
\right),\\
\hat{K}_{z}&=
\left(\begin{array}{cccc}
  0& 0 & -i  \\
0 & 0 & 0 \\
  i  & 0 & 0\\
  \end{array}
\right).
\end{aligned}
\end{eqnarray}
They satisfy the commutation relations
\begin{eqnarray}
[\hat{K}_{x},\hat{K}_{y}]=i\hat{K}_{z},~ [\hat{K}_{y},\hat{K}_{z}]=i\hat{K}_{x},~[\hat{K}_{z},\hat{K}_{x}]=i\hat{K}_{y}.
\end{eqnarray}
The fact that Hamiltonian (\ref{Hamilton}) is written as the sum of three SU(2) operators, means it addresses the SU(2) algebraic structure~\cite{Chen-PRA2012-STA}.

For the two enantiomers of chiral molecules, the overall phases in the cyclic three-level systems under consideration differ by $\pi$~\cite{Lee-PRL2022-ESST}. For convenience, we specify that the signs before $\Omega_{x}$ and $\Omega_{z}$ are equal for the two enantiomers, while the sign before $\Omega_{y}$ is opposite, as shown in Fig.~\ref{fig1}.

Therefore, the Hamiltonians of the cyclic three-level systems for the two enantiomers in the basis \{$|m\rangle_{L}$\} and \{$|m\rangle_{R}\}~(m=1,2,3)$ can be described as
\begin{eqnarray}
\label{H3}
\hat{H}^{L,R}(t)=\Omega_{x}(t)\hat{K}^{L,R}_{x}\pm\Omega_{y}(t)\hat{K}^{L,R}_{y}+\Omega_{z}(t)\hat{K}^{L,R}_{z}.
\end{eqnarray}
Here, the indices $L$ and $R$ [which correspond, respectively, to the signs $+$ and $-$ in the right side of Eq.~(\ref{H3})], denote the left- and right-handed chiral molecules, respectively. $\hat{K}_{j}^{Q}(j=x,y,z, Q=L,R)$ is just $\hat{K}_{j}$ in Eq.~(\ref{SU2}) for the two enantiomers.
In this work, when referring to left- or right-handed chiral molecules, we will add the index. When there is no index, we refer to general molecules.

\section{Invariant dynamics}

Shortcut to adiabaticity (STA) is a fast route to accelerate a slow adiabatic process by controlling the parameters of a system~\cite{Odelin-RMP2019-STA}, while keeping the same initial and final states as that in the adiabatic passage. A motivation to apply the STA technique is to manipulate the quantum system on timescales shorter than decoherence times.
There are two main STA techniques that have been proposed theoretically and implemented experimentally to inversely engineer the time-dependent Hamiltonian of a quantum system for accelerating slow adiabatic process~\cite{Chen-PRA2011-STA}. One is the counterdiabatic driving method with adding an auxiliary field in a reference Hamiltonian to cancel the nonadiabatic coupling, where the dynamics follows exactly the  adiabatic passage defined by the reference Hamiltonian~\cite{Chen-PRA2011-STA,Berry-JPA2009-STA,Chen-PRL2010-STA2}.
The other one is the invariant-based inverse engineering method, which is based on the Lewis-Riesenfeld invariant that carries the eigenstates of a system from the initial state to the desired final state~\cite{Chen-PRA2011-STA}, with keeping the same initial and final states as those in the adiabatic passage, but without following the adiabatic passage at the intermediate time instants~\cite{Chen-PRL2010-STA,Chen-PRA2011-STA}.
In what follows, we focus on how to use the latter STA technique to achieve the ESST of chiral molecules.

Commonly a Lewis-Riesenfeld invariant for a Hamiltonian $\hat{H}(t)$ is a Hermitian operator $\hat{I}(t)$ that satisfies~\cite{Lewis-JMP1969}
\begin{eqnarray}
\label{invariant}
\frac{d\hat{I}(t)}{dt}\equiv\frac{\partial \hat{I}(t)}{\partial t}-i[\hat{I}(t),\hat{H}(t)]=0,
\end{eqnarray}
so that its eigenvalues remain constant in time.
According to the Lewis-Riesenfeld theory~\cite{Lewis-JMP1969,Chen-PRL2010-STA,Chen-PRA2011-STA}, if $\{|\phi_{n}(t)\rangle\}$ is a set of orthogonal eigenstates of the invariant $\hat{I}(t)$, the solution to the time-dependent Sch\"{o}rdinger equation 
can be constructed as $|\Psi(t)\rangle=\sum_{n}c_{n}e^{i\alpha_{n}(t)}|\phi_{n}(t)\rangle$,  with $c_{n}$ being a time-independent coefficient.
Here $\alpha_{n}(t)=\int_{0}^{t}\langle\phi_{n}(t')|[i\partial_{t'}-\hat{H}(t')]|\phi_{n}(t')\rangle dt'$ is the Lewis-Riesenfeld phase~\cite{Lewis-JMP1969,Chen-PRL2010-STA,Chen-PRA2011-STA}.

In general, $\hat{H}(t)$ does not commute with the invariant $\hat{I}(t)$ at all time. We only require the invariant and the Hamiltonian to
commute at the initial and final time instants, i.e.,
$[\hat{H}(0),\hat{I}(0)]=0$ and $[\hat{H}(\tau),\hat{I}(\tau)]=0$~\cite{Chen-PRL2010-STA,Chen-PRA2011-STA,Chen-PRA2012-STA,Odelin-RMP2019-STA}. The eigenstates of the Hamiltonian and the invariant coincide at the initial and final time instants but may be different at the intermediate time. This leaves large freedom to choose how the state evolves in the intermediate time. We can use Eq.~(\ref{invariant}) to find the Hamiltonian~(\ref{Hamilton}) that drives such a designed evolution of a given state in the cyclic three-level system. Moreover, we consider, respectively, the evolutions of the left- and right-handed chiral molecules with cyclic three-level structures by invariant-based inverse engineering of the Rabi frequencies (equivalently the amplitude of the electromagnetic fields). By choosing appropriate Rabi frequencies, the fast ESST can be achieved by transferring the two enantiomers from their ground states to different-energy final states through their corresponding eigenstates of invariants, following their enantio-selective STA paths.

\subsection{Invariant dynamics for the left-handed chiral molecules}

We first consider the state transfer of the left-handed chiral molecules with the cyclic three-level structures by the invariant-based inverse engineering. Since $\hat{H}^{L}(t)$ in Eq.~(\ref{H3}) possesses the SU(2) algebraic structure, the corresponding invariant $\hat{I}^{L}(t)$ can be given as~\cite{Chen-PRA2012-STA}
\begin{eqnarray}
\label{10}
\hat{I}^{L}=&&\frac{\Omega_{0}}{2}(\cos\gamma\sin\beta\cdot \hat{K}^{L}_{x}+\cos\gamma\cos\beta\cdot \hat{K}^{L}_{y}+\sin\gamma\cdot \hat{K}^{L}_{z}) \nonumber\\
=&&\frac{\Omega_{0}}{2}\left(\begin{array}{cccc}
0&\cos\gamma\sin\beta &-i\sin\gamma\\
\cos\gamma\sin\beta & 0 & \cos\gamma\cos\beta\\
i\sin\gamma & \cos\gamma\cos\beta & 0 \\
  \end{array}
\right)_{L}
\end{eqnarray}
in the basis $\{|1\rangle_{L}, |2\rangle_{L}, |3\rangle_{L}\}$. Here, $\Omega_{0}$ is an arbitrary constant with unit of frequency, and the time-dependent auxiliary parameters $\gamma$ and $\beta$ satisfy the equations
\begin{eqnarray}
\label{Omegaleft}
\begin{aligned}
\dot{\gamma}&=\Omega_{x}\cos\beta-\Omega_{y}\sin\beta,\\
\dot{\beta}&=(\Omega_{x}\sin\beta+\Omega_{y}\cos\beta)\tan\gamma-\Omega_{z}.
\end{aligned}
\end{eqnarray}

The eigenstates of the invariant $\hat{I}^{L}(t)$, which satisfy $\hat{I}^{L}(t)|\phi_{n}(t)\rangle_{L}=\lambda_{n}^{L}|\phi_{n}(t)\rangle_{L}~(n=0,\pm)$, are
\begin{eqnarray}
|\phi_{0}\rangle_{L}&=&
\left(\begin{array}{cccc}
\cos\gamma\cos\beta \\
-i\sin\gamma \\
-\cos\gamma\sin\beta\\
  \end{array}
\right)_{L}, \label{8} \\
|\phi_{\pm}\rangle_{L}&=&\frac{1}{\sqrt{2}}
\left(\begin{array}{cccc}
\sin\gamma\cos\beta\pm i\sin\beta  \\
i\cos\gamma \\
-\sin\gamma\sin\beta\pm i\cos\beta\\
  \end{array}
\right)_{L}  \label{88}
\end{eqnarray}
with the corresponding (time-independent) eigenvalues $\lambda^{L}_{0}=0$ and $\lambda^{L}_{\pm}=\pm \Omega_{0}$. In this case, the Lewis-Riesenfeld phases are $\alpha^{L}_{0}(t)=0$, and $\alpha^{L}_{\pm}(t)=\mp\int_{0}^{t}[\dot{\beta}(t')\sin\beta(t')+\Omega_{x}(t')\sin\beta(t')\cos\gamma(t')+\Omega_{y}(t')\cos\beta(t')\cos\gamma(t')+\Omega_{z}(t')\sin\gamma(t')]dt'$.

Here, we take $\Omega_{x}(t)=\Omega_{z}(t)$ for simplicity. By using Eq.~(\ref{Omegaleft}), we have
\begin{eqnarray}
\begin{aligned}
\label{Omegadesignleft}
\Omega_{x}&=\Omega_{z}=\frac{\dot{\beta}\sin\beta+\dot{\gamma}\cos\beta\tan\gamma}{\tan\gamma-\sin\beta},\\
\Omega_{y}&=\frac{\dot{\beta}\cos\beta+\dot{\gamma}(1-\tan\gamma\sin\beta)}{\tan\gamma-\sin\beta}.
\end{aligned}
\end{eqnarray}
Once the appropriate boundary conditions for $\gamma$ and $\beta$ are fixed, one can insert a polynomial function to determine $\Omega_{x}$, $\Omega_{y}$, and $\Omega_{z}$. Our task is to design the Hamiltonian $\hat{H}^{L}(t)$ to drive the initial state $|1\rangle_{L}$ to the final state $|3\rangle_{L}$ (up to a phase factor) along the invariant eigenstate $|\phi_{0}(t)\rangle_{L}$ in a given time $\tau$. Therefore, based on the invariant eigenstate $|\phi_{0}(0)\rangle_{L}= (1,0,0)^{T}_{L}=|1\rangle_{L}$ at the initial instant time and $|\phi_{0}(\tau)\rangle_{L}=(0,0,-1)^{T}_{L}=-|3\rangle_{L}$ at the final instant time $\tau$, the boundary conditions for $\gamma$ and $\beta$ can be given as
\begin{eqnarray}
\begin{aligned}
\label{11}
&\gamma(0)=0,~\beta(0)=0,\\
&\gamma(\tau)=0,~\beta(\tau)=\frac{\pi}{2}.
\end{aligned}
\end{eqnarray}
On one hand, one needs to impose the boundary conditions to make $\hat{H}^{L}(t)$ and $\hat{I}^{L}(t)$ commute at $t = 0$ and $t=\tau$ so that they have common eigenstates at these time instants.
On the other hand, one requires the Rabi frequencies to vanish at the initial and final time instants
to make the electromagnetic fields turn on and off smoothly. These requirements further imply the additional boundary conditions
\begin{eqnarray}
\begin{aligned}
\dot{\gamma}(0)=0,~\dot{\beta}(0)=0,\\
\dot{\gamma}(\tau)=0,~\dot{\beta}(\tau)=0.
\end{aligned}
\end{eqnarray}
There are many interpolating functions consistent with the boundary conditions at the initial and final time instants.
With these boundary conditions, we can simply choose
\begin{eqnarray}
\label{gamaleft}
\gamma(t)=0,~
\beta(t)=\frac{3\pi}{2\tau^{2}}t^{2}-\frac{\pi}{\tau^{3}}t^{3}+\eta.
\end{eqnarray}
Here the small value $\eta$ is set to avoid the infinite values of the Rabi frequencies at the initial time instant.
Thus the designed Rabi frequencies in Eq.~(\ref{Omegadesignleft}) reduce to
\begin{eqnarray}
\begin{aligned}
\Omega_{x}&=\Omega_{z}=\frac{3\pi t}{\tau^{2}}\left(\frac{t}{\tau}-1\right),\\
\Omega_{y}&=\frac{3\pi t}{\tau^{2}}\left(\frac{t}{\tau}-1\right)\cot\left(\frac{3\pi}{2\tau^{2}}t^{2}-\frac{\pi}{\tau^{3}}t^{3}+\eta\right).\\
\end{aligned}
\label{Rabileft}
\end{eqnarray}
\begin{figure}[htbp]
\centering
\includegraphics[width=250pt]{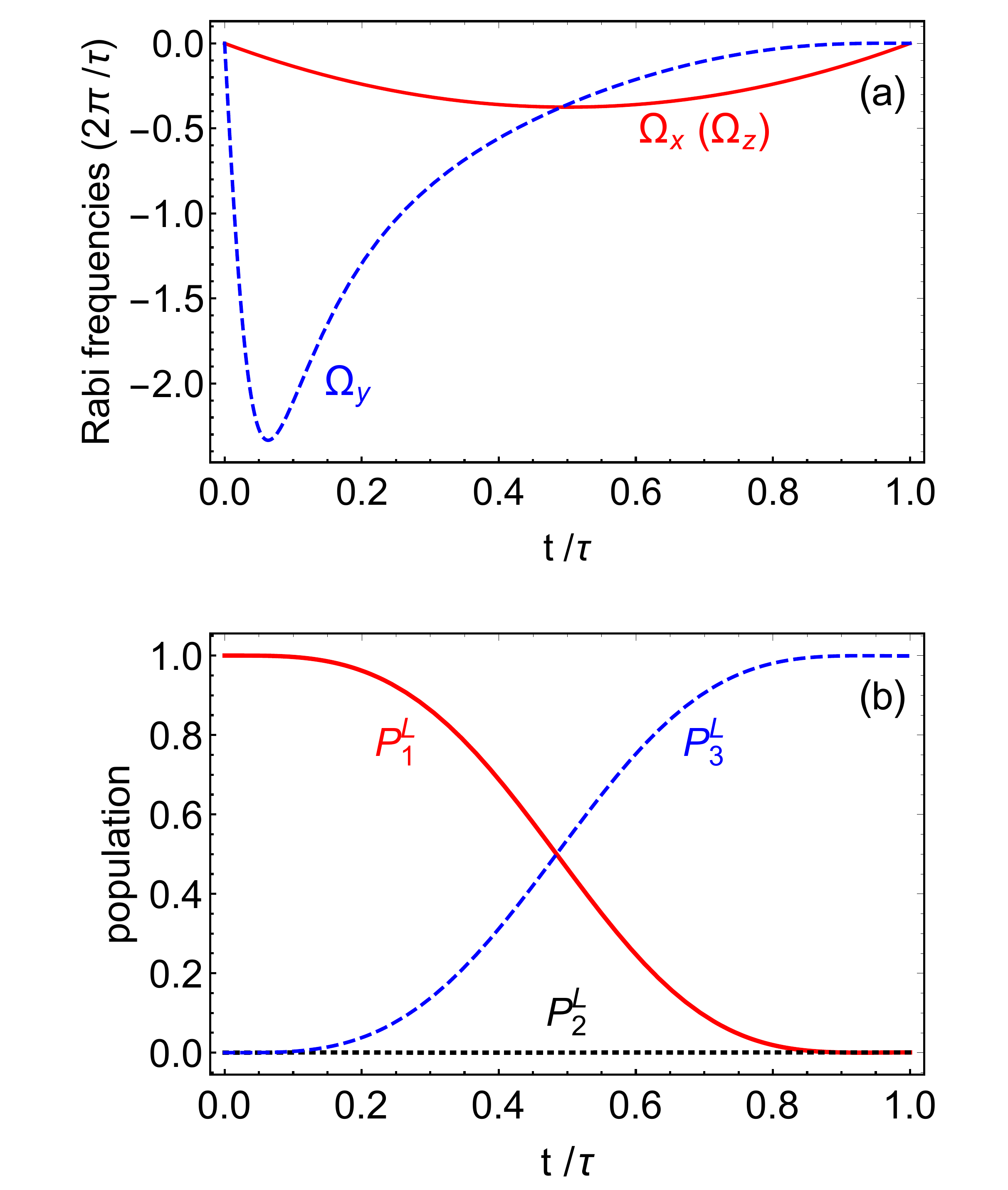}
\caption{ (Color online) (a) The designed Rabi frequencies for the left-handed chiral molecules with $\Omega_{x}=\Omega_{z}$ (red solid line) and $\Omega_{y}$ (blue dashed line) given in Eq.~(\ref{Rabileft}). (b) Time evolution of corresponding populations in $|1\rangle_{L}$~(red solid line),~$|2\rangle_{L}$ (black dotted line), and~$|3\rangle_{L}$~(blue dashed line) for the left-handed chiral molecules with the initial state $|1\rangle_{L}$. Here $\eta=0.02$.}
\label{fig2}
\end{figure}

Fig.~\ref{fig2} shows the designed Rabi frequencies for the left-handed chiral molecules and corresponding evolution of the populations in the states $|m\rangle_{L}$ ($m=1,2,3$) for the initial state $|\Psi(0)\rangle_{L}=|1\rangle_{L}$. In the ideal condition (i.e. the case of $\eta=0$), the  left-handed chiral molecules will evolve from the initial state $|1\rangle_{L}~(=|\phi_{0}(0)\rangle_{L})$ to the final target state $-|3\rangle_{L}$ (up to a phase factor), along the invariant eigenstate $|\phi_{0}(t)\rangle_{L}$. For the case of small value $\eta=0.02$ as shown in Fig.~\ref{fig2}(b), the initial state $|1\rangle_{L}\approx|\phi_{0}(0)\rangle_{L}$, thus the populations in the initial state $|1\rangle_{L}$ with $P_{1}^{L}(0)=1$ are finally transferred approximately to that in the target state $|3\rangle_{L}$ with probability $P_{3}^{L}(\tau)=0.9991$ for the left-handed chiral molecules. Correspondingly, $P_{2}^{L}(0)=0=P_{3}^{L}(0)$, $P_{1}^{L}(\tau)=0.0005$, and $P_{2}^{L}(\tau)=0.0004$.

\subsection{Invariant dynamics for the right-handed chiral molecules}

Then we consider the state transfer of the right-handed chiral molecules with the cyclic three-level structures by the invariant-based inverse engineering.
Since the Hamiltonian $\hat{H}^{R}(t)$ in Eq.~(\ref{H3}) of the right-handed chiral molecules has the same SU(2) algebraic structure as $\hat{H}^{L}(t)$ of the left-handed ones, similarly the invariant $\hat{I}^{R}(t)$ can be given in the basis $\{|1\rangle_{R}, |2\rangle_{R}, |3\rangle_{R}\}$ as the form
\begin{eqnarray}
\label{15}
\hat{I}^{R}&&=\frac{\Omega_{0}}{2}(\cos\xi\sin\chi\cdot \hat{K}^{R}_{x}+\cos\xi\cos\chi\cdot \hat{K}^{R}_{y}+\sin\xi\cdot \hat{K}^{R}_{z})\nonumber\\
&&=\frac{\Omega_{0}}{2}\left(\begin{array}{cccc}
0&\cos\xi\sin\chi &-i\sin\xi\\
\cos\xi\sin\chi & 0 & \cos\xi\cos\chi\\
i\sin\xi & \cos\xi\cos\chi & 0 \\
  \end{array}
\right)_{R}.
\end{eqnarray}
Here the time-dependent auxiliary parameters $\xi(t)$ and $\chi(t)$ satisfy the equations
\begin{eqnarray}
\label{Omegaright}
\begin{aligned}
\dot{\xi}&=\Omega_{x}\cos\chi+\Omega_{y}\sin\chi,\\
\dot{\chi}&=(\Omega_{x}\sin\chi-\Omega_{y}\cos\chi)\tan\xi-\Omega_{z}.
\end{aligned}
\end{eqnarray}
The eigenstates of the invariant $\hat{I}^{R}(t)$, which satisfy $\hat{I}^{R}(t)|\phi_{n}(t)\rangle_{R}=\lambda^{R}_{n}|\phi_{n}(t)\rangle_{R}~(n=0,\pm)$, are
\begin{eqnarray}
|\phi_{0}\rangle_{R}&=&
\left(\begin{array}{cccc}
\cos\xi\cos\chi\\
-i\sin\xi \\
-\cos\xi\sin\chi\\
  \end{array}
\right)_{R}, \\
|\phi_{\pm}\rangle_{R}&=&\frac{1}{\sqrt{2}}
\left(\begin{array}{cccc}
\sin\xi\cos\chi\pm i\sin\chi  \\
i\cos\xi \\
-\sin\xi\sin\chi\pm i\cos\chi\\
  \end{array}
\right)_{R}
\end{eqnarray}
with the corresponding eigenvalues $\lambda^{R}_{0}=0$ and $\lambda^{R}_{\pm}=\pm\Omega_{0}$. Here the Lewis-Riesenfeld phase is $\alpha^{R}_{0}(t)=0$, and $\alpha^{R}_{\pm}(t)=\mp\int_{0}^{t}[\dot{\chi}(t')\sin\chi(t')+\Omega_{x}(t')\sin\chi(t')\cos\xi(t')-\Omega_{y}(t')\cos\chi(t')\cos\xi(t')+\Omega_{z}(t')\sin\xi(t')]dt'$.

Here we still take $\Omega_{x}=\Omega_{z}$ for simplicity. According to Eq.~(\ref{Omegaright}), we have
\begin{eqnarray}
\begin{aligned}
\label{Omegadesignright}
\Omega_{x}&=\Omega_{z}=\frac{\dot{\chi}\sin\chi+\dot{\xi}\cos\chi\tan\xi}{\tan\xi-\sin\chi},\\
\Omega_{y}&=\frac{\dot{\chi}\cos\chi+\dot{\xi}(1-\tan\xi\sin\chi)}{\sin\chi-\tan\xi}.
\end{aligned}
\end{eqnarray}
Similar to the case of the left-handed chiral molecules in the above subsection, once the functions $\chi$ and $\xi$ are fixed, we can construct $\Omega_{x}$, $\Omega_{y}$, and $\Omega_{z}$ and thus the Hamiltonian $H^{R}(t)$ can be determined.
Here we aim to design
the Hamiltonian $\hat{H}^{R}(t)$ to make the system evolve from the initial state $|1\rangle_{R}$ to the finial state $|2\rangle_{R}$ (up to a phase factor) along the invariant eigenstate $|\phi_{0}(t)\rangle_{R}$ in a given time $\tau$. Therefore, based on the invariant eigenstate $|\phi_{0}(0)\rangle_{R}= (1,0,0)^{T}_{R}=|1\rangle_{R}$ at the initial time instant and $|\phi_{0}(\tau)\rangle_{R}=(0,-i,0)^{T}_{R}=-i|2\rangle_{R}$ at the final time instant $\tau$, the boundary conditions for $\xi$ and $\chi$ can be given as
\begin{eqnarray}
\begin{aligned}
\label{19}
\xi(0)=0,~\chi(0)=0,~\xi(\tau)=-\frac{\pi}{2}.
\end{aligned}
\end{eqnarray}

\begin{figure}[htbp]
\centering
\includegraphics[width=250pt]{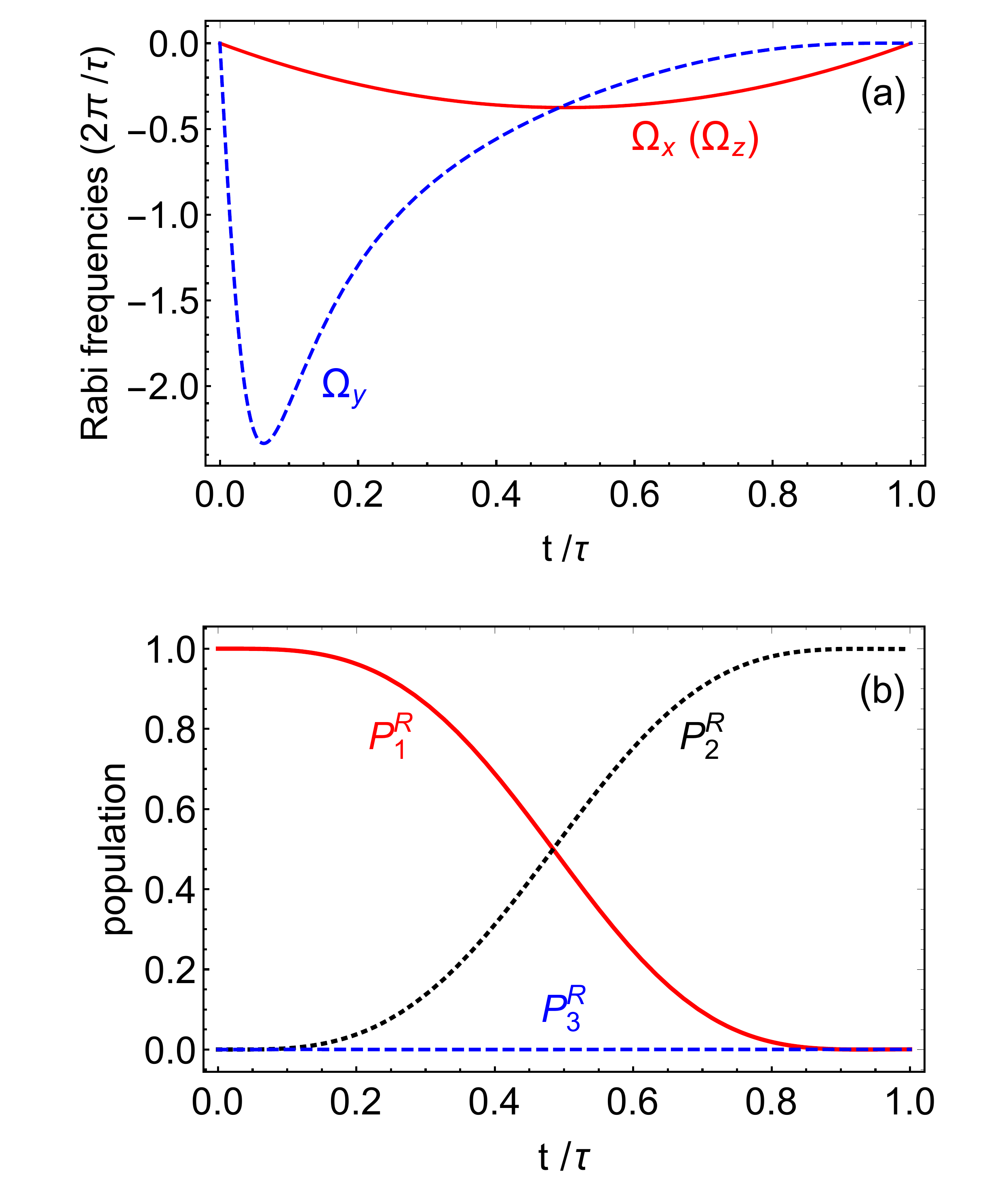}
\caption{(Color online) (a) The designed Rabi frequencies for the right-handed chiral molecules with $\Omega_{x}=\Omega_{z}$ (red solid line) and $\Omega_{y}$ (blue dashed line) given in Eq.~(\ref{Rabiright}). (b) Time evolution of corresponding populations in $|1\rangle_{R}$ (red solid line), $|2\rangle_{R}$ (black dotted line), and $|3\rangle_{R}$ (blue dashed line) for the right-handed chiral molecules with the initial state $|1\rangle_{R}$. Here $\eta^{\prime}=-0.02$.}
\label{fig3}
\end{figure}
Similarly, we set $\hat{H}^{R}(t)$ and $\hat{I}^{R}(t)$ commute at the initial and final time instants (so that they have the same eigenstates at these time instants) and make the electromagnetic fields (equivalently the Rabi frequencies) turn on and off smoothly for the right-handed chiral molecules. Thus, the additional boundary conditions for $\xi(t)$ and $\chi(t)$ can be given as
\begin{eqnarray}
\begin{aligned}
\dot{\xi}(0)=0,~\dot{\chi}(0)=0,\\
\dot{\xi}(\tau)=0,~\dot{\chi}(\tau)=0.
\end{aligned}
\end{eqnarray}
Consistent with these boundary conditions, we can choose
\begin{eqnarray}
\label{xiright}
\chi(t)=0,~
\xi(t)=-\frac{3 \pi}{2\tau^{2}}t^{2}+\frac{\pi}{\tau^{3}}t^{3}+\eta'.
\end{eqnarray}
Here the small value $\eta'$ is set to avoid the infinite values of the Rabi frequencies at the initial time instant.
Thus the designed Rabi frequencies in Eq.~(\ref{Omegadesignright}) reduce to
\begin{eqnarray}
\begin{aligned}
\label{Rabiright}
\Omega_{x}&=\Omega_{z}=\frac{3\pi t}{\tau^{2}}\left(\frac{t}{\tau}-1\right),\\
\Omega_{y}&=\frac{3\pi t}{\tau^{2}}\left(\frac{t}{\tau}-1\right)\cot\left(\frac{3\pi}{2\tau^{2}}t^{2}-\frac{\pi}{\tau^{3}}t^{3}-\eta'\right).\\
\end{aligned}
\end{eqnarray}

\begin{figure}[htbp]
\centering
\includegraphics[width=250pt]{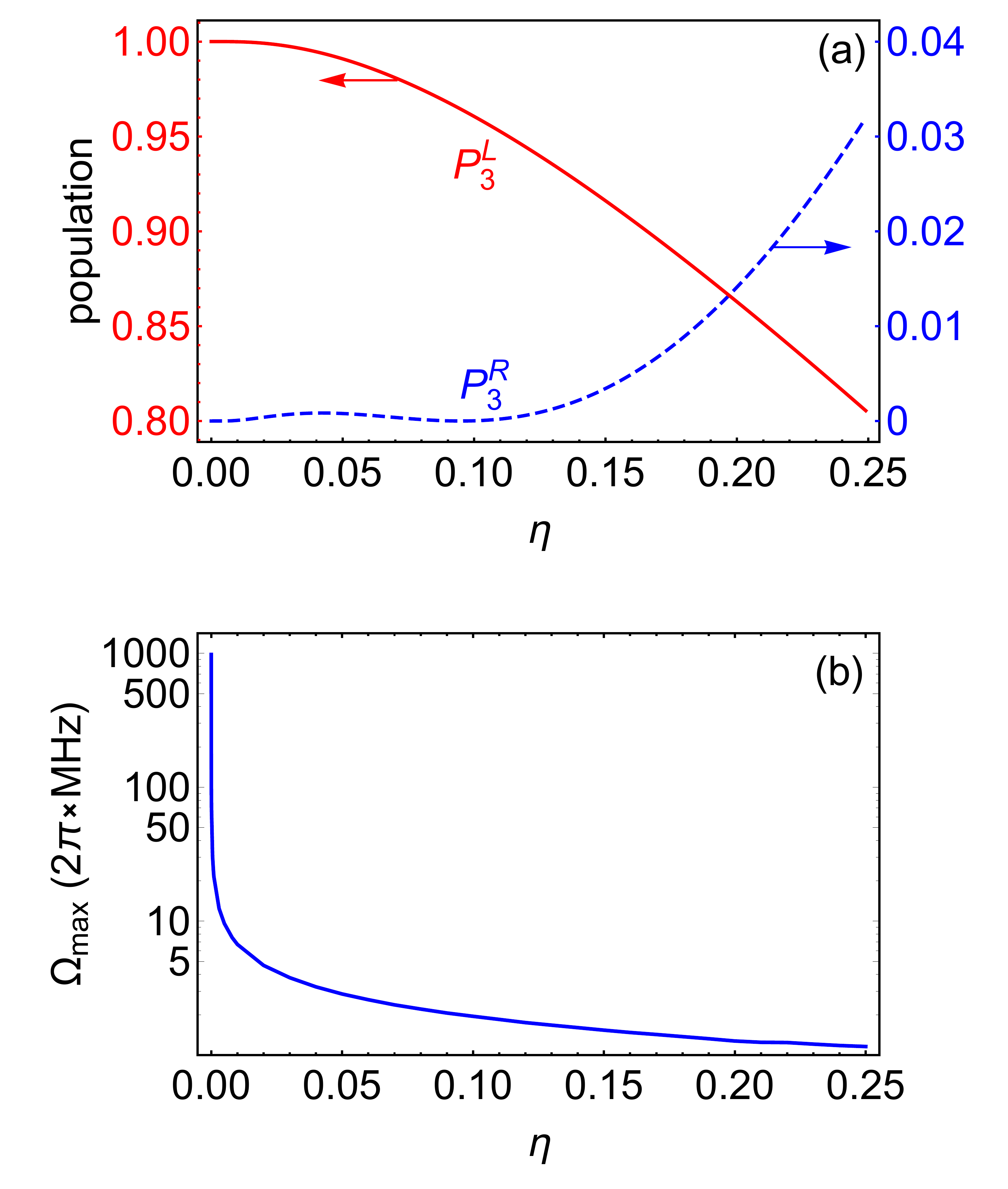}
\caption{(Color online) (a) The corresponding  populations in $|3\rangle_{L}$ (red solid line) and $|3\rangle_{R}$ (blue dashed line) at the final time versus the small value $\eta$. The initial states are $|1\rangle_{L,R}$.
(b) The maximum absolute value of the Rabi frequencies $\Omega_{\mathrm{max}}$ versus the small value $\eta$ with $\tau=0.5\,\mu$s.}
\label{fig4}
\end{figure}

Fig.~\ref{fig3} shows the designed Rabi frequencies of the right-handed chiral molecules
and corresponding evolution of the populations in the states $|m\rangle_{R}$ ($m=1,2,3$) for the initial state {$|\Psi(0)\rangle_{R}=|1\rangle_{R}$.

In the ideal condition (i.e. the case of $\eta'=0$), the right-handed chiral molecules will evolve from the initial state $|1\rangle_{R}~ (=|\phi_{0}(0)\rangle_{R})$ to the final target state $-i|2\rangle_{L}$ (up to a phase factor), along the invariant eigenstate $|\phi_{0}(t)\rangle_{R}$. When we set the small value $\eta'=-0.02$ as shown in Fig.~\ref{fig3}(b), the initial state $|1\rangle_{R}\approx|\phi_{0}(0)\rangle_{R}$, thus  the populations in the initial state $|1\rangle_{R}$ with  $P_{1}^{R}(0)=1$ are finally transferred approximately to that in the target state $|2\rangle_{R}$ with $P_{2}^{R}(\tau)=0.9991$ for the right-handed chiral molecules. Correspondingly, $P_{2}^{R}(0)=0=P_{3}^{R}(0)$, $P_{1}^{R}(\tau)=0.0005$, and $P_{3}^{R}(\tau)=0.0004$.

\subsection{Achieving the fast enantio-specific state transfer}

So far we have designed the desired evolution for the left- and right-handed chiral molecules of the cyclic three-level systems via the STA technique with invariant-based inverse engineering in the above two subsections, respectively. By comparing Eq.~(\ref{Rabileft}) with Eq.~(\ref{Rabiright}), it can be found that the two groups of designed Rabi frequencies for the two enantiomers are exactly the same when $\eta=-\eta'$. This means that the two enantiomers are driven by the same three electromagnetic fields indeed. In this case, the left-handed chiral molecule begins with $|1\rangle_{L}$ and terminates approximately at $-|3\rangle_{L}$, almost along the invariant eigenstate $|\phi_{0}(t)\rangle_{L}$, while the right-handed chiral molecule begins with $|1\rangle_{R}$ and terminates approximately at $-i|2\rangle_{R}$, almost along the invariant eigenstate $|\phi_{0}(t)\rangle_{R}$ simultaneously. As also shown in Fig.~\ref{fig2} and Fig.~\ref{fig3}, the left- and right-handed chiral molecules prepared in the same-energy initial states evolves (approximately) to the different-energy final states via the different enantio-selective STA processes of invariant-based inverse engineering, driven by the same electromagnetic fields. Thus, the fast ESST via enantio-selective STA is achieved (approximately).

In the above ESST method via the enantio-selective STA with invariant-based inverse engineering, the enantiomeric excess of the ESST can be defined as~\cite{Ye-PRA2019-discrimination,Zhang-JPB2020-ESST}
\begin{eqnarray}
\epsilon\equiv\Big|\frac{P^{L}_{3}(\tau)-P^{R}_{3}(\tau)}{P^{L}_{3}(\tau)+P^{R}_{3}(\tau)}\Big|.
\end{eqnarray}
Although the small values $\eta$ and $\eta^{\prime}$ (e.g. $\eta=-\eta^{\prime}=0.02$) have been introduced to avoid the infinite $\Omega_{y}$ at the initial time instant, we still obtain a highly efficient ESST with enantiomeric excess $\epsilon=99.92\%$ at the final time instant (with most of left-chiral molecule staying in $|3\rangle_{L}$
and very few of the right-chiral molecule staying in the same-energy state $|3\rangle_{R}$, as shown in Fig.~\ref{fig2} and Fig.~\ref{fig3}).

In general, the final populations  are effected by the small value $\eta$ (or $\eta'$) and are independent of the parameter $\tau$. As shown in Fig.~\ref{fig4}(a), the population of the target state $|3\rangle_{L}$ can be further decreased by increasing  the small value $\eta$, while the population of  the other target state $|3\rangle_{R}$  would be commonly increased by increasing the small value $\eta$. Therefore, it is possible to achieve a better enantiomeric excess with relatively  small value $\eta$. According to Eq.~(\ref{Rabileft}) and Eq.~(\ref{Rabiright})}, decreasing the small amount $\eta$ (or $\eta^{\prime}$) implies the tradeoff of requiring larger Rabi frequencies and laser intensities~\cite{Chen-PRA2012-STA}.
Here we define $\Omega_{\mathrm{max}}$=Max$\{|\Omega_{x}(t)|,|\Omega_{y}(t)|, |\Omega_{z}(t)|\}$ as the maximum absolute value of the Rabi frequencies during the whole evolution process. 
As shown in Fig.~\ref{fig4}(b), the maximum absolute value of the Rabi frequencies increase dramatically when decreasing the small value  $\eta$.

In experiments, the typical Rabi frequencies for the transitions of chiral molecules are about $2\pi\times10$\,MHz~\cite{Patterson-PRL2013-discrimination,
Eibenberger-PRL2017-Experimental,Perez-Angew2017-Experimental}. That means the evolution time can be shortened to be 0.5\,$\mu$s for the experimentally available Rabi frequencies. Thus, the decoherence effects (typically being about $5\sim6\,\mu$s)~\cite{Patterson-Nature2013-discrimination,Eibenberger-PRL2017-Experimental} will  become negligable. This is the advantage of our ESST method since it allows to manipulate the quantum system on the timescales much shorter than the typical decoherence time.

Note that in the previous ESST method via STA~\cite{Vitanov-PRL2019-ESST}, an auxiliary counterdiabatic field has been applied. It works as a shortcut to adiabaticity for canceling the nonadiabatic coupling and induces perfect population transfer between the states $|1\rangle_{L}$ and $|3\rangle_{L}$ for the left-handed chiral molecules. Simultaneously, it also acts oppositely for strengthening the nonadiabatic coupling for the right-handed chiral molecules and the population transfer between the states $|1\rangle_{R}$ and $|3\rangle_{R}$ is canceled completely. Therefore, under such an ESST process, the left-handed chiral molecule begins with $|1\rangle_{L}$ and terminates at $-|3\rangle_{L}$, following a STA path. But the right-handed chiral molecule is subject to a free evolution, instead of following the STA path. By contrast, in our ESST method via STA, the eigenstates of invariants for the two enantiomers define their corresponding enantio-selective STA paths. Thus, our ESST can be achieved with transferring the two enantiomers from their ground states to different-energy final states along their enantio-selective STA paths simultaneously, by choosing appropriate intensities of the three electromagnetic fields (that is, the Rabi frequencies).

\section{Conclusion}
In conclusion, we have proposed the fast ESST method of chiral molecules via the STA technique with invariant-based inverse engineering.  Based on the cyclic three-level systems, the ESST of chiral molecules can be achieved through enantio-selective STA paths: for the left- and right-handed chiral molecules prepared initially in their ground states, they will evolve (approximately) finally to the different-energy states almost along the eigenstates of the invariants within a short operation time simultaneously. Hence, our fast ESST method via STA with invariant-based inverse engineering has promising applications in discriminating molecular chirality and controlling the dynamics of chiral molecules.

\section*{ACKNOWLEDGMENTS}
This work was supported by the Natural Science Foundation of China (Grants No.~12074030, No.~12274107, and
No.~U1930402), National Science Foundation for Young Scientists of China (No.~12105011), and Beijing Institute of Technology Research Fund Program for Young Scholars.

\end{document}